\documentclass[a4]{revtex4}

\usepackage{amssymb,amsfonts,amsmath,mathrsfs,gensymb}
\usepackage{graphicx}
\usepackage[normalem]{ulem}
\usepackage{hyperref}

\newcommand{\ESCORE}{${\mathcal E}$SCORE }
\newcommand{\ERMSD}{${\mathcal E}$RMSD }
\newcommand{\ee}{{\cal E}}
\usepackage{color,soul}

\begin{document}

\title{A nucleobase-centered coarse-grained representation for structure prediction of RNA motifs.}

\author{%
Sim\'on Poblete\,$^{1,*}$,
Sandro Bottaro\,$^{1}$
and Giovanni Bussi\,$^1$%
\footnote{To whom correspondence should be addressed.
Email: spoblete@sissa.it,bussi@sissa.it}}

\address{%
$^{1}$
Scuola Internazionale Superiore di Studi Avanzati, \\
265, Via Bonomea I-34136 Trieste, Italy
}

\begin{abstract}
We introduce the SPlit-and-conQueR (SPQR) model, a coarse-grained representation of RNA designed for structure prediction and refinement. In our approach, the representation of a nucleotide consists of a point particle for the phosphate group and an anisotropic particle for the nucleoside. The interactions are, in principle, knowledge-based potentials inspired by the $\mathcal E$SCORE function, a base-centered scoring function. However, a special treatment is given to base-pairing interactions and certain geometrical conformations which are lost in a raw knowledge-base model. This results in a representation able to describe planar canonical and non-canonical base pairs and base-phosphate interactions and to distinguish sugar puckers and glycosidic torsion conformations.
The model is applied to the folding of several structures, including duplexes with internal loops of non-canonical base pairs, tetraloops, junctions and a pseudoknot. For the majority of these systems, experimental structures are correctly predicted at the level of individual contacts. We also propose a method for efficiently reintroducing atomistic detail from the coarse-grained representation.

\end{abstract}

\maketitle
\section{Introduction}

During the last decades, RNA has been found to be much more than a mere messenger and translator of the genetic information in the cell. Its enzymatic and regulatory function has been observed in a variety of cellular processes, confering it a major role in evolution and cellular metabolism \cite{doudna2002,lehman2010, mattick2004, serganov2013,morris2014}. For the thorough understanding of these functions, an insight on the three-dimensional structure of RNA molecules is of crucial importance. Nevertheless, the reliable prediction of the full structure of an RNA motif based uniquely on its sequence is still a challenging aim. 

RNA is dominantly composed by a mixture of the four most common nucleotides. Its alphabet is thus in principle significantly simpler than the one used by proteins. This, together with the simple rule governing Watson-Crick pairing, has suggested RNA folding and structure prediction to be a relatively easy task \cite{tinoco1999}. However, RNA structural complexity is considerable due to the large number of backbone conformations \cite{murray2003, richardson2008} and the rich variety of interactions \cite{leontis2001, leontis2002, zirbel2009} which play a substantial contribution in the stability of many biologically relevant structures, and blind structure prediction is difficult to perform \cite{puzzles1, puzzles2,puzzles3}.

In this respect, many computational approaches have been proposed to accomplish this goal during the last decades. All-atom Molecular Dynamics simulations \cite{sponer2014} would in principle allow to fold small motifs \cite{chen2013}. Nevertheless, they are usually limited by the large amount of computational resources required and the reliability of the force fields, which still need certain refinement to be trusted quantitatively \cite{cheatham2015, bottaro2016}. In another vein, bioinformatic assembly tools \cite{mcfold1, mcfold2, farna, rnabuilder, assemble, rna2d3d, rsim, rnalien, moderna, rnacomposer} can increase the sampling efficiency, but they also posess limitations coming from their energy functions, they do not necessarily provide information about the folding pathway and might fail when predicting motifs that are not already characterized experimentally and thus were not used in their training phase.
Coarse-grained (CG) models thus emerge as a suitable alternative, at the expenses of resolution and versatility \cite{ifoldrna, CGren, hireRNA, simrna, nast, yup, rnapps, nares2p, vfold, thirumalai, toprna, rnakb, ragtop, ernwin, jever, oxrna}. Relying on different representations and interactions, several models have been proposed for the study of thermodynamic and mechanical properties, and/or to efficiently sample the conformational space for structure prediction and refinement purposes. 
Depending on the properties to reproduce and the criteria of the authors, CG models focus on particular interactions and structural features. For example, many of them restrict their base-base interactions to stacking and canonical base pairs \cite{oxrna}, or might model hairpin loops as loose, non-interacting segments of RNA \cite{vfold}. However, more recent approaches have put emphasis on tertiary contacts by representing nucleobases as anisotropic objects, able to form directional interactions and non-canonical base pairs (e.g., \cite{CGren, simrna, hireRNA}), although their inclusion is still challenging, and in some cases, limited \cite{puzzles1, puzzles2, puzzles3}.

\enlargethispage{-65.1pt}

In this paper, we introduce the SPQR (SPlit and conQueR) model, a nucleotide-level coarse-grained representation developed for the accurate prediction of the secondary and tertiary structure of small RNA motifs. The mapping of the nucleobases is based on the \ESCORE function \cite{escore, notapaper}, a scoring function which focuses exclusively on the relative arrangement of bases comparing it with the one observed in a structural database. Additionally, sugar and phosphate groups are represented as a virtual site and a point particle, respectively (see Figure \ref{rep1}). The interactions between these elements are designed to sample the geometrical probability distributions obtained from a large set of experimental structures from a base-centered perspective. However, these distributions have been carefully partitioned, identifying several of their contributions and reweighting their corresponding interactions. In this manner, the model focuses on an accurate geometrical description of base-base interactions (planar canonical and non-canonical base pairs) and base-phosphate interactions, while it rescues the conformation of the glycosidic bond angle and the sugar pucker, which are important elements in many relevant motifs.

We show that these interactions suffice for describing a number of structures involving a variety of canonical as well as non-canonical base pairs, like duplexes, tetraloops, three-way junctions and a pseudoknot. We also show that our base-centered approach is useful for the insertion of atomistic details in the predicted structure in a consistent framework.

The paper is organized as follows: The model and the interactions are exposed in the Methods section, 
together with the simulation protocol used in the paper. In the next section, the results will be presented over several sets of structures. 
We will begin with a proof of concept of our method, to continue with a set of tetraloops, double strands (with and without internal loops),  
a pseudoknot, a subset of motifs already tested in the FARFAR protocol \cite{farfar} and finally, a small set of junctions. The details of the structures can be found in the SI, section 1.
We later present the results of the backmapping procedure, which consists in the reintroduction of the
 atomistic resolution in the predicted structures, applied two tetraloops. Finally, the paper finishes with the conclusion and discussing how to improve the results presented here.

\section{MATERIALS AND METHODS}

\subsection{Coarse-grained representation}
The representation of a nucleotide consists of two elements: a point particle for the phosphate group and a rigid, anisotropic particle for the nucleoside,
 as shown in Figure \ref{rep1}. The base is represented as a triplet of particles forming a triangle which determines its centroid and orientation, as defined
 in the \ESCORE function \cite{escore}. Meanwhile, a virtual site rigidly attached to the base represents the geometrical center of the sugar ring. 
Each nucleotide has a well defined sugar pucker (C2'-{\it endo} or C3'-{\it endo}) and a glycosidic bond state, which can be chosen between \emph{anti} and \emph{high-anti} conformations, and \emph{syn} for purines. 

\subsection{Interactions}
The energy function between two nucleotides $i$ and $j$ is defined as
\begin{equation}
U_{ij}=U_{ij}^{EV} + U_{ij}^{bp}+U_{ij}^{st}+U_{ij}^{bph} + U_{ij, \chi p}^{bb}
\end{equation}
which is a combination of excluded volume (EV), base-pairing (bp), stacking (st), base-phosphate (bph) and interactions along the backbone (bb). The latter depends both on the glycosidic bond angle and the sugar pucker, a conformation that will be referred as the $\chi p$ state of a nucleotide from now on. This energy term is given by
\begin{equation}
U_{ij,\chi p}^{bb}=  U^b_{ij, \chi p} + U^a_{ij, p} + \frac{1}{2}\left(U_{i, \chi p}^s+U_{j, \chi p}^s\right) ,   \qquad j=i+1
\end{equation}
where $U^a_{ij, p}$ is an energy function of the angle formed by the sugar-phosphate-sugar triplet of two consecutive nucleotides, and its functional form depends on the sugar pucker conformations of the nucleotides involved. $U^b_{ij, \chi p}$ is the interaction between nucleoside $i$ and the phosphate group of nucleotide $i+1$. In addition, the self term $U_{i, \chi p}^s$ is a sum of the energy between a nucleoside and its own phosphate group and a shift $\epsilon_i^{\chi p}$ which characterizes the glycosidic bond angle and sugar pucker conformations of the nucleotide. 

\begin{figure}[t]
\begin{center}
\includegraphics[width=8cm]{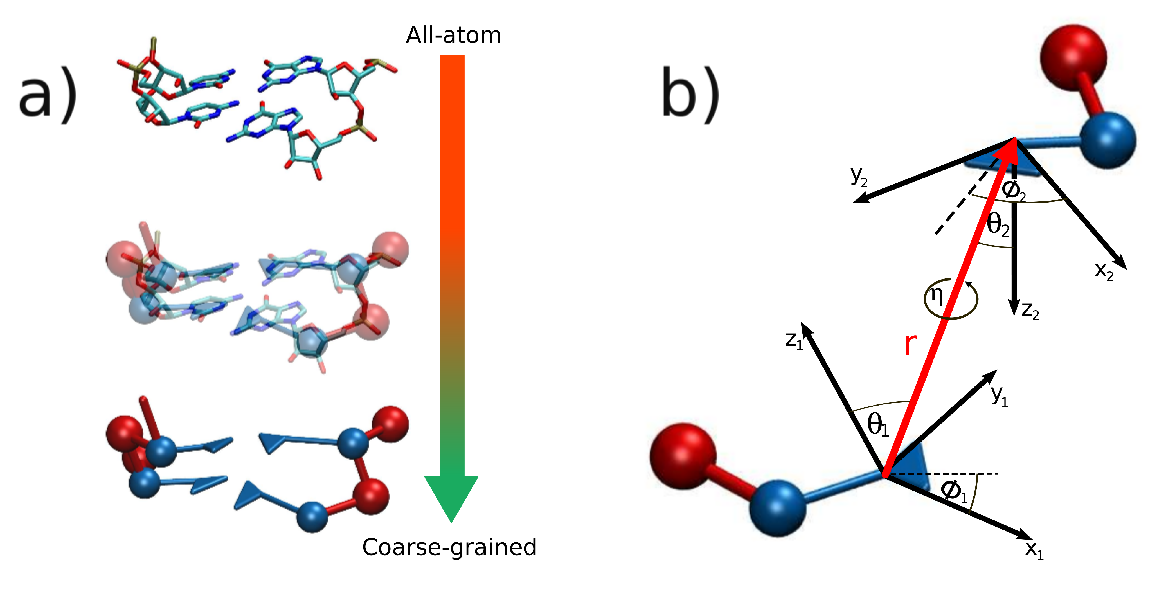}
\end{center}
\caption{a) Schematic representation of the mapping of four nucleotides with the nucleoside in blue and the phosphate in red. b) The definition of the coordinate system from the oriented base, for two nucleotides. Dashed lines represent the projection of the vector that joins both bases in the x-y plane of each nucleoside. Both base-base interactions and base-phosphate interactions along the backbone are defined with respect to this reference frame.}
\label{rep1}
\end{figure}

Each energy term is designed to reproduce a probability distribution between two or more types of particles. This distribution is sampled from all their occurrences in a non-redundant list of structures (see SI, section 1, \cite{non-red-list}). The histograms of the planar base-pairing region obtained are spanned by classifying the points according to the specific kind of interaction using the FR3D package \cite{fr3d}, and therefore, we rely on its definition of base pairing and hydrogen bonding. Thus, it is possible to associate a particular distribution for stacking interactions and non-canonical base pairs, as well as base-phosphate and backbone interactions. The interaction between two bases turns out to be the most complex due to the anisotropy of both particles, but also the most important one to describe the RNA structure. The probability distribution of the spatial configuration of two nucleosides depends on six coordinates, which are shown in Figure \ref{rep1}b. The \ESCORE reference frame deals only with the angles $\theta_i$, $\phi_i$ and the distance $r$. This allows to introduce the probability distribution function $P^\ee(r,\theta,\phi)$ based on these degrees of freedom. It is therefore reasonable to approximate the full probablity distribution as

\begin{align}
\label{sep-variable}
\nonumber P(r,\theta_1, \phi_1,& \theta_2, \phi_2, \eta) \approx \\ 
&r^2 P_\eta(\eta) \times \frac{P^\ee_1(r, \theta_1, \phi_1)P^\ee_2(r, \theta_2, \phi_2)}{H_r(r)}
\end{align}
where $P_\eta$ and $H_r$ are the probability distribution of the orientation $\eta$ and the histogram of the distance $r$, respectively. The remaining factors include the probability distribution of the distance $r$ and the corresponding Jacobian. 

\begin{figure}[t]
\begin{center}
\includegraphics[width=9cm]{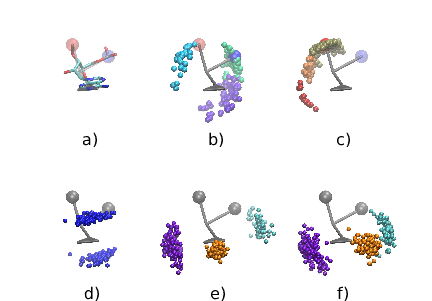}
\end{center}
\caption{Depiction of selected clouds of points found in structural database: a) Coarse-grained and atomistic representations of Adenine as found in a typical duplex. The own phosphate group and the one of the following nucleotide are colored in blue and red in the CG representation, respectively. b) Clouds of the nucleotide's own phosphate group; all clouds represent C3'-{\it endo} cloud under different $\chi$ conformations: {\it anti} (green), {\it high-anti} (purple) and {\it syn}(cyan). c) Clouds of the neghboring nucleotide's phosphate group position, $\chi$ in {\it anti} conformation with sugar in C3'-{\it endo} (gold) and in C2'-{\it endo} (orange). Also, the {\it syn} conformation is shown in red. 
d) Cloud of stacking points, e) cloud of positions of paired bases through sugar (purple), Watson-Crick (orange) and Hoogsteen (cyan) faces. f) represents the positions of the phosphate groups for base-phosphate interactions, with the same color nomenclature of e).
}
\label{rep2}
\end{figure}

Once done this approximation, the interaction potential is obtained by Boltzmann inversion, as
\begin{align}
\nonumber U(r,\theta_1, \phi_1,  &\theta_2, \phi_2, \eta) = \\
&-T_0\log P(r,\theta_1, \phi_1, \theta_2, \phi_2, \eta) + \epsilon
\label{energy}
\end{align}
where $T_0$ is a fictitious temperature. Here, we use the approximation of Eq. \ref{sep-variable} for the function $P$ and introduce a shift $\epsilon$, which corrects the arbitrary normalization of the probability distributions. This indetermination can be used to control the relative strength of the different interactions present in the system (see Figure \ref{rep2}). In a first instance, in the planar non-canonical base pairs, $\epsilon$ is chosen proportional to the number of hydrogen bonds present in the Leontis-Westhof tables (see \cite{leontis2002}), while the stacking interactions are adjusted with a fitting procedure over a series of structures (see section Parametrization). This approach is intended to distinguish the multiple interactions that emerge not only between different species, but also between the Watson-Crick, Hoogsteen and Sugar/CH4 faces. Stacking interactions, on the other side, depend on the orientation of the planes defined by the bases involved. This makes that $\epsilon$ depends on the kind of interaction, species and faces involved in the pairing or stacking.
For the base-phosphate interactions, the energy is given by Eq. \ref{energy} but using $P^\ee(r, \phi, \theta)$ as the probability distribution, due to the lack of structure of our phosphate group representation. In this case, the splitting of the probability distribution is also done according to the base faces involved. The position of the base with respect to the phosphate group in its local reference frame is also taken into account to form the base-phosphate interaction and avoid false positives (see SI, section 2).

The backbone interactions are designed in a similar way. For the interactions between a nucleoside and its neighboring phosphates along the backbone, Eq. \ref{energy} is used with its corresponding $P^\ee$ probability distribution, which is classified according to their backbone conformations using the Dangle software (http://kinemage.biochem.duke.edu/software/suitename.php). The obtained probability clouds are found to depend strongly on the sugar pucker and the glycosidic bond of the involved base (see Figure \ref{rep2}). In this representation, they emerge as a natural partition of the conformational space, which keeps certain resemblance with previous pucker-dependent virtual bonds representations of the backbone \cite{olson1975}. The sugar-phosphate-sugar angle shows also a clear dependency on these conformations (see SI, section 2). In these cases, the absence of a special distinction of the conformations greatly favors the most populated conformation which corresponds to the C3'-{\it endo} pucker for the sugar pucker and {\it anti} conformation of the glycosidic bond torsion $\chi$. 

The excluded volume interaction is also taken from the \ESCORE representation, which assigns to sugar, bases and phosphates an ellipsoidal geometry with specific parameters.

\subsection{Simulation protocol}
Random moves are proposed and accepted with a standard Monte Carlo Metropolis procedure \cite{frenkel}. In addition, each nucleotide can change 
its $\chi p$ state by displacing and rotating the nucleobase and remapping the sugar, as described in the SI, section 3. %

Our method, in its most general form, is intended to be applied on two steps: first, a search of the global minimum of the energy function, which can be performed by Simulated Annealing or Simulated Tempering, while keeping the $\chi p$ state of all the nucleotides fixed in the {\it anti} and C3'-{\it endo} states. These calculations will be referred as $\chi p_c$ simulations along the paper, to stress the constraint on these variables. Then, a shorter annealing procedure is run on a smaller set of nucleotides, typically a junction or an internal or hairpin loop, without any constraint on the $\chi p$ states. In this manner, for example, we anneal junctions and internal and hairpin loops by anchoring the molecule to a rigid vicinity, which can be determined from a previous $\chi p_c$ simulation or from a crystal structure. However, for larger structures such as duplexes and pseudoknots, we have opted for taking into account their natural flexibility during this refinement step. To this aim, we apply a soft restraint on the secondary structure during the refinement.

For the folding simulations, we used a simulated annealing protocol with 20 initial conditions. For the anchored tetraloops, each step of the annealing procedure consisted of $5\times 10^7$ 
Monte Carlo trials on each nucleotide, saving conformations every 5000 steps. These parameters had slight modifications in the rest of the systems. The annealing procedure started at temperature $T=15T_0$. $T$ was multiplied by a factor 0.75 \cite{anneal-prot} 
when the minimum energy of an annealing step did not decrease. Once it reached a value smaller than $ T_0$, the temperature was set
 to zero so as to minimize the energy of the resulting structure and run for a time equal to the one of an annealing step. This is in general more than enough to obtain a converged structure . In addition, we also performed Simulated Tempering simulations for the pseudoknot,
which is also implemented in our code. We used the method of \cite{parkpande} for estimating the initial values of the relative weights, with a maximum temperature of 12.5 $T_0$ and minimum of 0.5 $T_0$, using 12 temperatures separated by $\Delta T=T_0$.

\subsection{Parametrization}

We parametrize the set of $\epsilon$ shifts of stacking and base-phosphate interactions by folding a small set of structures. The base-pairing interactions are scaled according to the number of hydrogen bonds between two nitrogen or oxygen containing groups according to the Leontis-Westhof classification \cite{leontis2002}.
The energy scale of stacking interactions has been adjusted to obtain an initial value that will be refined in a posterior step. We start with a trial value strong enough to correctly fold the stem of the GCAA tetraloop. Later on, the backbone interactions are multiplied by a prefactor, which is required to obtain the correct arrangement of bases in the loop region. In a following step, we introduce the base-phosphate interactions, and adjust their strength in order to keep the formation of the stem stable. Later, the stacking interactions have been adjusted by folding a duplex which containts the UU stacking (PDB id: 255D), providing a lower bound to their strength. Thus, the stacking strengths are parametrized to distinguish between purines and pyrimidines (see SI, section 2).
The parameters obtained were later on refined by annealing a set of 46 internal and hairpin loops from the 1S72 ribosomal structure. On the resulting structures, the native one was not always obtained, and a number of alternative intermediate structures was often obtained as well. We determined the minimum energy structure for a large set of parameters of stacking (see SI, section 2), and calculated the INF score only for the stacking interactions. Thus, we chose the set of parameters which maximized this score over the entire training set. In a similar manner, the $\epsilon^{\chi p}$ was obtained by reweighting these decoys and maximizing the number of correctly predicted cases. The detailed values are reported in the SI, section 3, showing that after the parametrization, the {\it anti} glycosidic conformation with the C3'-{\it endo} sugar pucker is still the most energetically favorable one, although not as drastically as when no shift is applied.

\subsection{Backmapping}
Once the minimum energy coarse-grained structure is identified, we produce an all-atom prediction by performing steered-molecular dynamics simulations.
The MD simulation is performed using an atomistic description of RNA in explicit water (TIP3P water molecules \cite{tip3p}, Amber99 force field \cite{amber99} with parmbsc0 \cite{parmbsc0} and $\chi_{OL3}$ corrections \cite{chiol3}  ) in a truncated dodecahedral box with Na$^+$ counterions \cite{counterions}.
An external force proportional to the \ERMSD \cite{escore} with respect to the target structure is applied using the Gromacs code 4.6.7 \cite{gromacs1} in combination with PLUMED \cite{plumed}. This means that the bases are positioned according to the coarse-grained structure, while the backbone atoms simply adjust to this arrangement. The temperature was of 350K, while the pulling constant was of 500 kJ/nm$^2$.
The possibility to use the \ERMSD as a steered variable has been already discussed in \cite{bottaro2016} where it was used so as to
enforce the correct fold in stem-loop structures.

\subsection{Assessment of prediction quality}
We use the standard root-mean-square deviation after optimal superposition (RMSD \cite{kabsch-rmsd}) and the Interaction Network Fidelity (INF) to compare the predicted structures with their native counterpart. The INF is given by \cite{inf}
\begin{equation}
INF=\sqrt{\left(\frac{TP}{TP+FP}\right)\times\left(\frac{TP}{TP+FN}\right)}
\end{equation}
where $TP$ stands for the correctly predicted contacts while $FP$ and $FN$ are the false positive and false negative numbers, respectively. This coefficient is calculated separately for stacking (st), canonical pairs (wc), non-canonical pairs (nc) and base-phosphate (bph) pairs.
The RMSD was calculated by using the position of the sugar, phosphate and backmapping the C2, C4 and C6 atoms from the CG representation.

\section{Results}

In the following we discuss the results obtained using our model on a series
of benchmark systems for which the native structure is already known by means
of X-ray or NMR spectroscopy. %

As mentioned in the Methods section, the simulation protocol might be applied in two steps for large structures, while for 
small motifs it can be applied without any $\chi p_c$ constraint, but keeping the RNA motif anchored to its neighbors.
These choices will be specified in each subsection.

\subsection{Validation of the annealing procedure}

We have observed including the change of both glycosidic bond angle conformation and sugar pucker increases considerably the complexity of the conformational space and, consequently, the simulation time.
We thus perform the annealing simulations without restrictions in the $\chi p$ space
on a selected set of systems only, in order to
validate the two-step procedure. In particular, we perform a {\it de novo} prediction
of a duplex formed of the sequence CCCCGGGG (pdb:1RXB) and a hairpin of sequence GGGCGCAAGCCU (pdb: 1ZIH).
In both cases, the contacts and three-dimensional structure are nicely recovered.
Equivalent results were also produced using the two-step procedure, which evidences the robustness of our approach.

\subsection{Tetraloops}

\begin{figure}[t]
\begin{center}
\includegraphics[width=8cm]{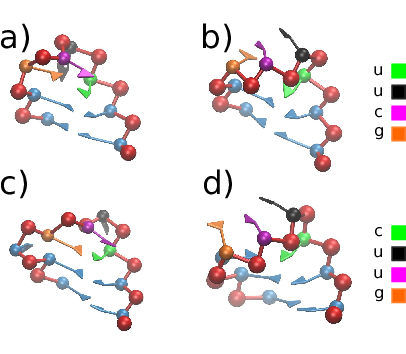}
\end{center}
\caption{a) UUCG tetraloop predicted by SPQR. b) UUCG tetraloop predicted by $\chi p_c$ annealing. c) CUUG tetraloop, closest structure to native and d) CUUG tetraloop predicted by $\chi p_c$ simulations.}
\label{flips}
\end{figure}

We have tested our approach on a set of tetraloops which include several GNRA occurrences and the UUCG and CUUG sequences. Details about the sequences and native structures can be found in the SI, section 4.
 The loop structures are optimized by a single round of Simulated Annealing while keeping them attached to a fixed stem 
obtained from the crystal structure.
 Only the tetraloops and the closing 
base pair are allowed to move during the simulation (6 flexible nucleotides). The total number of nucleotides
in the hairpin loops is 8 for GNRA and UUCG and 9 for CUUG. The change of the $\chi p$ state is allowed only on the four nucleotides in the tetraloop.
The results are shown in Table \ref{tetraloops}. We can observe the comparison between the systems with fixed $\chi p_c$ simulations.
In the case of GNRA and UUCG, the pucker and glycosidic bond angle conformations are correctly predicted, and the bases orient themselves
 accordingly leading to an excellent agreement.
Particularly, in UUCG, two bases are flipped : a purine (G9), by virtue of its glycosidic bond angle, and a pyrimidine
 (C8), due to its sugar pucker conformation, which is reconstructed by SPQR as shown in Figs. \ref{flips} a) and b). 
In the case of CUUG, the native structure was observed as the second most energetically favorable structure.
The energy difference between the native structure and the one predicted as most stable by SPQR was only of the order of the energy of a single hydrogen bond.
Nevertheless, the agreement is remarkable, reproducing both the loop region and a bulge as shown in Fig. \ref{flips} c), compared to its $\chi p_c$ counterpart in Fig. \ref{flips} d).

\begin{table}[b]
\caption{RMSD and INF scores of the predicted tetraloops.}
\label{tetraloops}
\begin{tabular*}{\columnwidth}{@{}lccccc@{}}
\toprule
Structure &RMSD (\AA) &  INF$^{ST}$ & INF$^{wc}$ & INF$^{nc}$ & INF$^{bph}$  \\
\\
\colrule
GCAA  & 1.4 &1   & 1   & 1 & 1\\
GAAA  & 1.7 &1   & 1   & 1 & 1\\
GAGA  & 1.2 &1   & 1   & 1 & 1\\
GCGA  & 1.8 &1   & 1   & 1 & 1\\
GGAA  & 2.1 &1   & 1   & 1 & 1\\
GGGA  & 1.3 &1   & 1   & 1 & 1\\
GUAA  & 1.3 &1   & 1   & 1 & 1\\
GUGA  & 1.5 &1   & 1   & 1 & 1\\
UUCG  & 0.9 &1   & 1   & 1 & 1\\
UUCG($\chi p_c$)&2.7 &0.86& 1   & 0 & 0\\
CUUG(s)&1.8 &0.94& 1   & - & 0\\
CUUG($\chi p_c$)&3.5 &0.7 & 0.81& 0 & -\\
\\
\botrule
\end{tabular*}%
{
  Results for the tetraloop simulations allowing the change of sugar pucker and $\chi$ angle conformation. Results forcing these parameters to the C3'-{\it endo} and {\it anti} conformation in all the nucleotides are denoted with ($\chi p_c$). Also, the second best result for CUUG is denoted by (s), and included for completeness.
}
\end{table}

\subsection{Double-stranded structures and pseudoknot}
We also tested our method on a set of double stranded-structures containing non-canonical pairs, as well as on a pseudoknot.
For these systems, the two-step procedure was applied as described in the Methods section, that is, we performed $\chi p_c$ simulations which are later refined imposing secondary structure restraints. The results before and after the refinement are shown in Table \ref{duplexes}. For the pseudoknot, all the canonical base pairs, with the exception of a pair of bulged bases, were correctly predicted.

\begin{table}[b]
\caption{RMSD and INF scores for the double-stranded structures and pseudoknot predictions.}
\label{duplexes}
\begin{tabular*}{\columnwidth}{@{}lccccc@{}}
\toprule
Structure &  RMSD (\AA) & INF$^{ST}$ & INF$^{wc}$ & INF$^{nc}$ &INF$^{bph}$\\
\\
\colrule
157D   & 1.6 & 0.98 & 1 & 1 &-\\
157D($\chi p_c$)& 1.5 & 0.98 & 1 & 1 &-\\
1DQH   & 2.3 & 0.93 & 1 & - &-\\
1DQH($\chi p_c$)& 2.1 & 0.93 &0.94&- &- \\
1KD5&4.4(3.6)& 0.81 & 1 & 0 &-(0)\\
1KD5($\chi p_c$)& 3.6 & 0.89 & 1 & 0 &-\\
1SA9   &2    &0.92  & 1 & 0 &-\\
1SA9($\chi p_c$)&2    &0.92  & 1 & 0 &- \\
205D   & 1.4 & 0.93 & 1 & 1 &-\\
205D($\chi p_c$)& 1.6 &0.95  & 1 & 1 &-\\
402D&3.3(1.4)&0.89  & 1 &  0(1)&-\\
402D($\chi p_c$)&1.4  &0.89  & 1 &1  &- \\
1I9X&    4.2 &0.85  &0.83&-&-\\
1I9X($\chi p_c$)& 4.5&0.85   &0.83&-&-\\
1L2X &   4&0.84&0.94&0.42&0\\
1L2X($\chi p_c$)&4.4&0.81&0.94&0.22&0\\
\botrule
\end{tabular*}%
{The values correspond to the scores of the minimum energy conformation found on each case. In parenthesis are the values of structures that were found with an energy practically undistinguishable to the minimum.}
\end{table}

\begin{figure}[t]
\begin{center}
\includegraphics[width=8cm]{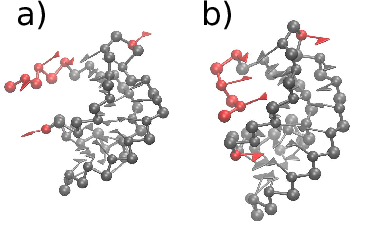}
\end{center}
\caption{a) Native pseudoknot 1L2X, b) folded after refinement. To facilitate the comparison between the two structures, the three initial nucleobases and two bulges are colored in red.}
\label{pseudok}
\end{figure}

Once the duplexes have been successfully folded with constraints, they are remarkably stable. In general, the results are not greatly affected by the additional degrees of freedom.
An unfavorable case is 1KD5, where a slightly less favorable conformation appears with a similar energy of the native with two puckers wrongly predicted. A similar case is observed in 402D, where an incorrect structure has practically the same energy as a better structure listed in the table. 
For the 1L2X pseudoknot structure we obtain a reasonable agreement without sampling glycosidic bond angle and pucker. However, it turns out that the refinement improves the INF corresponding to non-canonical pairs, and reduces the number of false positive base-phosphate interactions from 5 to 2.
For this system, the number of true positives is zero, so that INF is zero irrespectively of the number of false positives.
We also see that some bases can flip to {\it syn} state (the only cases on the present set), although they are unpaired or bulged. In addition, the first two unpaired nucleotides turn more flexible under the refinement. By neglecting them from the analysis, the RMSD improves to 3.3\AA. A depiction of this structure is found in Fig. \ref{pseudok}.

\subsection{Anchored internal loops and junctions}
We simulate a small subset of the FARFAR motifs \cite{farfar}. For a detailed description, see the SI, section 1. In this case, the motif is completely flexible while the nucleotides of the environment are frozen in space. This included the initial and terminal nucleotides of the strands involved, plus any interacting nucleotide according to FR3D or a nucleobase with a C4' atom separated by a distance smaller than 12\AA~ from any atom of the motif. The results are presented in Table \ref{farfar}.

\begin{table}[b]
\caption{RMSD and INF scores for subset of FARFAR motifs.}
\label{farfar}
\begin{tabular*}{\columnwidth}{@{}lccccc@{}}
\toprule
Structure &  RMSD(\AA) & INF$^{ST}$ & INF$^{wc}$ & INF$^{nc}$  & INF$^{bph}$\\
\\
\colrule
157D$_1$             &0.7  &1    &1      & 1   &-\\
157D$_1$($\chi p_c$) &0.7  &1    &1      & 1   &-\\
1D4R$_1$             &0.7  &0.94 &1      & 1   &-\\
1D4R$_1$($\chi p_c$) &0.6  &0.94 &1      &1  &-\\
1JJ2$_1$             &5    &0.55 &1      & 0           &0\\
1JJ2$_1$($\chi p_c$) &3.7  &0.62 &1   &	0.25  &0\\
1LNT$_1$    &1    & 0.96&1      &	0.89 &1\\
1LNT$_1$($\chi p_c$) &1.1	&0.96 &1      &	0.89 &1\\
1Q9A$_1$    &1.1  &0.91 &1      &	1    &1\\
1Q9A$_1$($\chi p_c$) &0.9 	&0.91 &1      &	1    &1\\
1U9S$_1$    &5.9  &0.22 &0.71   &	- &- \\
1U9S$_1$($\chi p_c$) &4.6  &0.67 &0.71   &	- &0 \\
2GDI$_1$    &0.9  & 1   &	1     &	- &-\\
2GDI$_1$($\chi p_c$) &2.3  &0.8  &	1     &	- &-\\
2OUI$_1$    &1.8  &0.75 &	1     &	0.5&-\\
2OIU$_1$($\chi p_c$) &2.1  &0.77 &	0.87  &	0  &-\\
2R8S$_1$    &1    &0.88 &	1     &1   &-\\
2R8S$_1$($\chi p_c$) &1    &0.88 & 1     &1   &-\\
2R8S$_3$    &3.6  &0.53 &	1     &	0.29&-\\
2R8S$_3$($\chi p_c$) &5.5  &0.5  &	1     &	0.29&-\\
\botrule
\end{tabular*}%
{}
\end{table}

The results are reasonable, showing that in many cases, when the $\chi p_c$ simulations produce good enough results, the additional degrees of freedom introduced by the $\chi p$ refinement do not compromise the structure. 

On average, the $\chi p$ refinement improves slightly the values of the INF for non-canonical base pairs. Nevertheless, it makes possible the formation of certain motifs which are technically forbidden without the additional freedom of the sugar pucker and glycosidic bond angle as in 2GDI (see SI, section 5). In addition, base flips seem not to cause a major problem; in fact, the only case they appear spuriously is in unpaired bases of 1JJ2.
The INF values obtained here are comparable to the values obtained with SimRNA for the same motifs, although the authors have reported a smaller RMSD.
Finally, Table \ref{junctions} shows the results for the junctions.

\begin{table}[b]
\caption{RMSD and INF scores of the set of junctions.}
\label{junctions}
\begin{tabular*}{\columnwidth}{@{}lccccc@{}}
\toprule
Structure &RMSD (\AA) &  INF$^{ST}$ & INF$^{wc}$ & INF$^{nc}$ & INF$^{bph}$  \\
\\
\colrule
1GID$_2$    & 7.9&0.47    &1       &0.63&0.71     \\
1GID$_2$($\chi p_c$) &3    &0.55	&1	 &0.71  &   0.35  \\
2QBZ$_3$   & 8.4&	0.30&	0.58  &0& 0 \\
2QBZ$_3$($\chi p_c$)& 6.5& 0.41&   0.87	 &0   &    0\\
3R4F$_2$ & 3.1&	0.59&	1	&-     &-\\
3R4F$_2$($\chi p_c$)&2.8  &0.6  &1	&-     &-\\
4P8Z$_2$ & 2.2   &0.47 &	1&-&	0\\
4P8Z$_2$($\chi p_c$)&2.9  &0.47 &1	&-  &   0\\
4P9R$_2$ & 3.3&	0.57&	1	&- &   -\\
4P9R$_2$($\chi p_c$)& 2.7	  &0.52 &1&- &  - \\
\\
\botrule
\end{tabular*}%

\end{table}

Here the results are in general worse than in the previous cases. Although the $\chi p$ refinement does not improve the results when they are already bad, it does improve them when the $\chi p_c$ predictions are good. Taken together, these results suggest that in order to be reliably used to predict junctions, the model should likely be further refined.

\subsection{Backmapping}
\begin{figure}[t]
\begin{center}
\includegraphics[width=8cm]{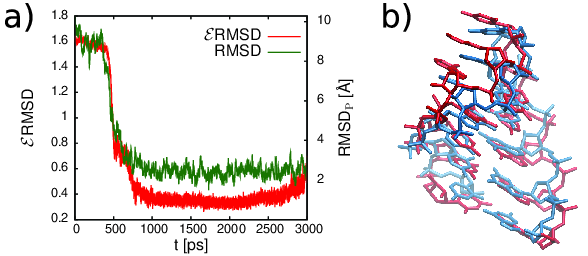}
\end{center}
\caption{a)\ERMSD and phosphate RMSD as a function of time for backmapping of 1ZIH. b) Comparison of GCAA tetraloop(1ZIH) at full-atom resolution for native (blue) and annealed (red).} 
\label{bm-all}
\end{figure}

The backmapping procedure has been applied over two representative tetraloops (GCAA, PDB id:1ZIH and UUCG, PDB id: 2KOC), using six initial conditions starting from a free strand and MD simulations of 3 ns. The pulling force is proportional to the \ERMSD between the bases of the atomistic structure and the lowest-energy CG structure found in the annealing procedure and thus effectively steers the atomistic model towards the structure predicted using the CG model.
We stress  that since this \ERMSD calculation is done with respect to the predicted model, its evaluation does not require any knowledge
of the native structure.
The \ERMSD converges quickly, as seen in Fig. \ref{bm-all}, reaching a value of 0.4 or lower when the folding is successful.
The RMSD of the phosphates also converges in the same time scale.
This result is not straightforward, since the \ERMSD only measures the relative nucleobase arrangement.
 By choosing the lowest \ERMSD structures of the simulations, we find an all-atom RMSD (without hydrogen atoms) from native of  2.3 \AA~ for 1ZIH and 1.9 \AA~ for 2KOC. 

The backmapping procedure is not always successful. In some cases, the base-flipping, like the one of G8 in the UUCG tetraloop, is not observed. This is an indication that the steering procedure was too fast, not letting the atomistic model
enough time to relax. Nevertheless, the \ERMSD of these structures is above 0.6, which makes these instances easy to detect
without any knowledge of the native structure.

By taking the 90 lowest \ERMSD structures, that is the atomistic structures that better approximate the predicted CG model,  we have analyzed the glycosidic bond torsion and sugar puckers in the tetraloop region. The sugar pucker was identified with MolProbity \cite{molprob}. Structures with not classified backbone conformations were discarded in the analysis. We observe that for GCAA, in 84\% of the cases the full tetraloop has the right glycosidic bond conformation in all its nucleotides, from which 92\% had the right conformation in all the sugar puckers at the same time. On the other side, UUCG shows 76\% of right glycosidic conformations, from which 85\% of them have the right puckers. Note that these loops are not static in solution, so it is not unexpected to observe a variety of rotameric states \cite{bottaro2017}. 

\section{DISCUSSION AND CONCLUSION}

We proposed a coarse-grained model with a hybrid parametrization for folding small RNA motifs. The model pays special attention to the non-canonical base pairs, base phosphate interactions, sugar puckers and glycosidic bond torsions, which are the key elements that determine, from our study, the geometrical arrangement of nucleotides from the perspective of a base.
It is important to observe that whereas the shape of the distributions associated to each interaction type is obtained from
the employed non-redundant list of structures, the strength of the base-pairs interactions follows the classification presented in \cite{leontis2002}. This gives more relevance to certain base-pairs that are not very frequent but could have a significant energetic contribution. It also depends of both faces of the bases when they are confronted. Therefore, the correction also compensates the different normalizations which cannot be captured by the usual approximations that consider, in a first instance, the interactions as in an structureless liquid \cite{skolnick2000}. An equivalent procedure is also used to partition the space according to the $\chi p$ values.

The procedure used for the parametrization of our model makes it a hybrid in between models which are based on purely statistical potentials (e.g., \cite{simrna,farna}) and
methods that are trained based on the energetics of the relevant interactions (e.g., \cite{thirumalai}).

We also have opted for leaving out the pseudotorsional space \cite{pyle-pstorsions} interactions. Although there is evidence of the correlation of these variables and the total RMSD of certain structures, it is not clear whether they are indeed the cause of the form of each loop, and they are not able to distinguish between sugar puckers. We observe that our representation sufficed in many cases for the correct prediction of several motifs.
This suggests that the incorporation of additional contributions to the potential is likely not necessary at this time.
We also noticed that incorporating additional potentials based on statistics from structural databases may shift the predicted structures to more frequent conformations in the reference
database, e.g., A-form duplexes, if not properly orthogonalized to the already present contributions.

Although the parametrization we present is not unique and could be further refined, the simple choice we have used in this paper has been enough for folding a number of structures which include canonical and non-canonical pairs. Even more, we have shown that the inclusion of puckers and glycosidic bond torsions allows for an improvement in the predicted structures in cases where rare conformations ({\it syn} and C2'-{\it endo}) are present in the native structure, without significantly destabilizing the most common conformations. Also, crucial base-phosphate interactions have been successfully recovered. Considering the complexity of the interactions, it is also remarkable that the annealing from random conformations is able to reproduce both duplexes and hairpins with good accuracy of bulges and non-canonical pairs.

The blind prediction tests give results that are comparable to those obtained with SimRNA \cite{simrna}. However, it should be mentioned that an intrinsic advantage of our model is that it allows base flipping to be explicitly modeled changing both sugar puckers and glycosidic bond angle. We thus think that SPQR could be optimally used side-by-side with other modeling tools that do not take these degrees of freedom explicitly into account.

We have also implemented a backmapping procedure able to introduce atomistic detail on the candidate structures. Although the procedure is designed to only enforce the position and orientation of the bases, it is able to backmap reasonably well the whole structure, with a good agreement of the sugar puckers and glycosidic bond torsion conformations in most cases. This allows our protocol to be used for structure prediction at
atomistic resolution.
We recall that atomistic molecular dynamics is still impaired by
sampling issues and by the accuracy of force fields and it is not able to predict correctly such a variety of motifs.
On the opposite side, structural bioinformatics tools are typically designed so as to quickly model or fold larger structures
with an accuracy that does not allow individual contacts to be reliably predicted. Our model is just optimally suited in the middle,
and might be useful for refining structures obtained with other tools so as to bring them at atomistic resolution.

The coarse-grained simulations were performed using an in-house Monte Carlo code, which can be downloaded at \url{http://github.com/srnas/spqr}.
The folding of a 12-nt hairpin may take around 6 hours in a desktop PC. Depending on the number of resources available, this procedure can be parallelized for a better exploration of the folding space, although both the code and the annealing procedure can be improved. 
The typical cost of a backmapping simulation with our procedure is 4 hours on twenty processors for
a 12-nt hairpin.
This time might seem large, considering that most of the available coarse-grained models allow for the atomistic representation
to be constructed in a single step, typically by assembling fragments that are compatible with the coarse-grained representation.
However, the timescale of the backmapping simulation itself is comparable or even shorter than that of the structure prediction
using the coarse-grained model, so that the total simulated time is less than doubled if
only the best prediction is to be refined. In addition,
if one is willing to use the resulting structure as an input for an atomistic simulation, our procedure has the
advantage of producing a fully solvated and equilibrated structure. Finally, the backmapping procedure itself,
being based on a force field with a reasonable accuracy, can help in identifying nucleotides that have been wrongly placed
by the coarse-grained simulation. The similarity between the timescales of the coarse-grained simulation and
of the backmapping procedure naturally calls for a sinergic simulation protocol where both representations
are simultaneously evolved, that will be the subject for further investigation.

In this paper we presented a novel protocol for RNA structure prediction and refinement.
The employed parametrization and the separation of the variables along the backbone
were shown to be the key ingredients for the formation of tetraloops belonging to three families, as well as for the
correct prediction of a number of hairpins, duplexes, internal loops, junctions and a pseudoknot, featuring multiple non-canonical interactions.
Future work will contemplate a further refinement of the base-base interactions and an extensive validation on a larger number of
motifs taken from structural databases.

\section{ACKNOWLEDGEMENTS}
The research
leading to these results has received funding from the European Research
Council under the European Union\textquoteright{}s Seventh Framework
Programme (FP/2007-2013) / ERC Grant Agreement n. 306662, S-RNA-S.

\subsubsection{Conflict of interest statement.} None declared.


\newpage


\begin{thebibliography}{4}


\bibitem{lehman2010}
Lehman, N. (2010)
RNA in evolution.
\textit{Wiley Interdiscip. Rev. RNA}, \textbf{1}, 202--213.

\bibitem{doudna2002}
Doudna, J. A. and Cech, T. R. (2002)
The chemical repertoire of natural ribozymes.
\textit{Nature}, \textbf{418}, 222--228.

\bibitem{mattick2004}
Mattick, J. S. (2004)
RNA regulation: a new genetics?
\textit{Nat. Rev. Genet.}, \textbf{5}, 316--323.

\bibitem{serganov2013}
Serganov, A. and Nudler, E. (2013)
A decade of riboswitches.
\textit{Cell}, \textbf{152}, 17--24.

\bibitem{morris2014}
Morris, K. V. and Mattick, J. S. (2014)
The rise of regulatory RNA.
\textit{Nat. Rev. Genet.}, \textbf{15}, 423--437.

\bibitem{tinoco1999}
Tinoco, I. Jr. and Bustamante, C. (1999)
How RNA folds.
\textit{J. Mol. Biol.}, \textbf{293}, 271--281.

\bibitem{murray2003}
Murray, L.W., Arendall III, W.B., Richardson, D.C., and
Richardson, J.S. (2003)
RNA backbone is rotameric.
\textit{Proc. Natl. Acad. Sci. U.S.A.}, \textbf{100}, 13904--13909.

\bibitem{richardson2008}
Richardson, J. S., Schneider, B., Murray, L. W., Kapral, G. J., Immormino, R. M., Headd, J. J., Richardson, D. C., Ham, D., Hershkovits, E., Williams et al. (2008)
RNA backbone: Consensus all-angle conformers and modular string nomenclature (an RNA Ontology Consortium contribution).
\textit{RNA}, \textbf{14}, 465--481.

\bibitem{leontis2001}
Leontis, N. B. and Westhof, E. (2001)
Geometric nomenclature and classification of RNA base pairs.
\textit{RNA}, \textbf{7}, 499--512.

\bibitem{leontis2002}
Leontis, N. B., Stombaugh, J. and Westhof, E. (2002)
The non‐Watson–Crick base pairs and their associated isostericity matrices.
\textit{Nucl. Acids Res.}, \textbf{30}, 3497--3531.

\bibitem{zirbel2009}
Zirbel, C. L., \v{S}poner, J. E., \v{S}poner, J., Stombaugh, J., and Leontis, N. B. (2009)
Classification and energetics of the base-phosphate interactions in RNA.
\textit{Nucl. Acids Res.}, \textbf{37}, 4898--4918.

\bibitem{puzzles1}
Cruz, J. A., Blanchet, M. F., Boniecki, M., Bujnicki, J. M., Chen, S. J., Cao, S., Das, R., Ding, F., Dokholyan, N. V., Flores, S. C., et al.(2012)
RNA-Puzzles: A CASP-like evaluation of RNA three-dimensional structure prediction.
\textit{RNA}, \textbf{14}, 610--625.

\bibitem{puzzles2}
Miao, Z., Adamiak, R. W., Blanchet, M.-F., Boniecki, M., Bujnicki, J. M., Chen, S.-J., Cheng, C., Chojnowski, G., Chou, F.-C., Cordero, P., et al.(2015)
RNA-puzzles round II: assessment of RNA structure prediction programs applied to three large RNA structures.
\textit{RNA}, \textbf{21}, 1066--1084.

\bibitem{puzzles3}
  Miao, Z., Adamiak, R. W., Antczak, M., Batey, R. T. , Becka, A. J., Besiada, M., Boniecki, M., Bujnicki, J. M., Chen, S. J., Cheng, C. Y. et al. (2017)
  RNA-Puzzles Round III: 3D RNA structure prediction of five riboswitches and one ribozyme.
  \textit{RNA}, \textbf{23} (5), 655--672.

\bibitem{sponer2014}
\v{S}poner, J., Banas, P., Jurecka, P., Zgarbova, M., Kuhrova, P., Havrila, M., Krepl, M., Stadlbauer, P., Otyepka, M. (2014)
Molecular Dynamics Simulations of Nucleic Acids. From Tetranucleotides to the Ribosome.
\textit{J. Phys. Chem. Lett}, \textbf{5}, 1771--1782.


\bibitem{chen2013}
Chen, A. A. and Garc\'ia, A. E. (2013)
High-resolution reversible folding of hyperstable RNA tetraloops using molecular dynamics simulations.
\textit{Proc. Natl. Acad. Sci. U.S.A.}, \textbf{110}, 16820--16825.





\bibitem{cheatham2015}
Bergonzo, C., Henriksen, N. M., Roe, D. R. and Cheatham, T. E., 3rd (2015)
Highly Sampled Tetranucleotide and Tetraloop Motifs Enable
Evaluation of Common Rna Force Fields.
\textit{RNA}, \textbf{21}, 1578--1590.



\bibitem{bottaro2016}
Bottaro, S., Bana\v{s}, P., \v{S}poner, J. and Bussi, G. (2016)
Free energy landscape of GAGA and UUCG RNA tetraloops.
\textit{J. Phys. Chem. Lett.}, \textbf{7}, 4032--4038.



\bibitem{mcfold1}
Parisien, M. and Major, F. (2008)
The MC-Fold and MC-Sym pipeline infers RNA structure from sequence data.
\textit{Nature}, \textbf{452}, 51--55.



\bibitem{mcfold2}
Reinharz, V., Major, F. and Waldispuhl, J. (2012)
Towards 3D structure prediction of large RNA molecules: an integer programming framework to insert local 3D motifs in RNA secondary structure.
\textit{Bioinformatics}, \textbf{28}, i207--i214.



\bibitem{farna}
Das, R. and Baker, D. (2007)
Automated de novo prediction of native-like RNA tertiary structures.
\textit{Proc. Natl. Acad. Sci. U.S.A.}, \textbf{104}, 14664--14669.



\bibitem{rnabuilder}
Flores, S. C., Wan, Y., Russel, R and, Altman, R. B. (2010)
Predicting RNA structure by multiple template homology modeling.
\textit{Pac. Symp. Biocomput.}, \textbf{2010}, 216--227.



\bibitem{assemble}
Jossinet, F., Ludwig, T. E., and Westhof, E. (2010)
Aseemble: An interactive graphical tool to analyze and build RNA architectures at the 2D and 3D levels.
\textit{Bioinformatics}, \textbf{26}, 2057--2059.



\bibitem{rna2d3d}
Martinez, H. M., Maizel, J. V. Jr. and Shapiro, B. A. (2008)
RNA2D3D: A program for generating, viewing, and comparing 3-dimensional models of RNA.
\textit{J. Biomol. Struct. Dyn.}, \textbf{25}, 669--683.


\bibitem{rsim}
Bida, J. P. and Maher L. J., III (2012)
Improved prediction of RNA tertiary structure with insights into native state dynamics.
\textit{Bioinformatics}, \textbf{18}, 385--393.



\bibitem{rnalien}
Eggenhofer, F. , Hofacker, I. L. and H\"oner zu Siederdissen, C. (2016)
RNAlien - Unsupervised RNA family model construction.
\textit{Nucl. Acids Res.}, \textbf{44}, 8433--8441.



\bibitem{moderna}
Rother, M., Rother K., Puton, T. and Bujnicki, J. M. (2011)
ModeRNA: A tool for comparative modeling of RNA 3D structure.
\textit{Nucl. Acids Res.}, \textbf{36}, 1227--1236.

\bibitem{rnacomposer}
Popenda, M., Szachniuk, M., Antczak, M., Purzycka, K. J., Lukasiak, P., Bartol, N., Blazewicz, J. and Adamiak, R. W. (2012)
Atumated 3D structure composition for large RNAs.
\textit{Nucl. Acids Res.}, \textbf{40}, e112.



\bibitem{ifoldrna}
Sharma, S., Ding, F. and Dokholyan, N. V. (2008)
iFoldRNA: three-dimensional RNA structure prediction and folding.
\textit{Bioinformatics}, \textbf{24}, 1951--1952.

\bibitem{CGren}
Xia, Z., Gardner, D. P. , Gutell, R. R. and Ren, P. (2010)
Coarse-grained model for simulation of RNA three-dimensional structures.
\textit{J. Phys. Chem. B}, \textbf{114}, 13497--13506.


\bibitem{hireRNA}
Pasquali, S. and Derreumaux, P. (2010)
HiRE-RNA: A High Resolution Coarse-Grained Energy Model for RNA.
\textit{J. Phys. Chem. B.}, \textbf{114}, 11957--11966.




\bibitem{simrna}
Boniecki, M. J., Lach, G., Dawson, W. K., Tomala, K., Lukasz, P., Soltysinski, T., Rother, K. M. and Bujnicki, J. M. (2015)
SimRNA: a coarse-grained method for RNA folding simulations and 3D structure prediction.
\textit{Nucl. Acids Res.}, \textbf{44}, e63.


\bibitem{nast}
Jonikas, M. A., Radmer, R. J., Laederach, A., Das, R., Pearlman, S., Herschlag, D. and Altman, R. B. (2009)
Coarse-grained modeling of large RNA molecules with knowledge-based potentials and structural filters.
\textit{RNA}, \textbf{15}, 189--199.


\bibitem{yup}
Tan, R. K. Z., Petrov, A. S. and Harvey, S. C. (2006)
YUP: A molecular simulation program for coarse-grained and multi-scaled models.
\textit{J. Chem. Theory Comput.}, \textbf{2}, 529--540.


%
%
%
%
\bibitem{rnapps}
  Li, J., Zhang, J., Wang, J., Li, W. and Wang, W. (2016)
  Structure prediction of RNA loops with a probabilistic approach.
  \textit{PLoS Comput. Biol.}, \textbf{12}, e1005032.

\bibitem{nares2p}
Liwo, A., Baranowski, M., Czaplewski, C., Golas, E., He, Y., Jagiela, D., Krupa, P., Maciejczyk, M., Makowski, M., Mozolewska, M. A. et al. (2014)
A unified coarse-grained model of biological macromolecules based on mean-field multipole–multipole interactions
\textit{J. Mol. Model.}, \textbf{20}, 2306.


\bibitem{vfold}
Xu, X., Zhao, P. and Chen, S.-J. (2014)
Vfold: A web server for RNA structure and folding thermodynamics prediction.
\textit{PLoS ONE}, \textbf{9}, e107504.


\bibitem{thirumalai}
Denesyuk, N. A. and Thirumalai, D. (2013)
Coarse-Grained Model for Predicting RNA Folding Thermodynamics.
\textit{J. Phys. Chem. B}, \textbf{117}, 4901--4911.


\bibitem{toprna}
Mustoe, A. M., Al-Hashimi, H. M. and Brroks, C. K., III (2014)
Coarse grained models reveal essential contributions of topological constraints to the conformational free energy of RNA bulges.
\textit{J. Phys. Chem. B}, \textbf{118}, 2615--2627.


\bibitem{rnakb}
Bernauer, J., Huang, X., Sim, A. Y. L and Levitt, M. (2011)
Fully differentiable coarse-grained and all-atom knowledge-based potentials for RNA structure evaluation.
\textit{RNA}, \textbf{17}, 1066--1075.


\bibitem{ragtop}
Kim, N., Zahran, M. and Schlick, T. (2015)
Computational prediction of riboswitch tertiary structures including pseudoknots by RAGTOP: a hierarchical graph sampling approach.
\textit{Methods Enzymol.}, \textbf{553}, 115--135.


\bibitem{ernwin}
Kerpedjiev, P. , H\"oner zu Siederdissen, C. and Hofacker, I. L. (2015)
Predicting RNA 3D structure using a coarse-grain helix-centered model.
\textit{RNA}, \textbf{21}, 1110--1121.


\bibitem{jever}
Jost, D. and Everaers, R. (2010)
Prediction of RNA multiloop and pseudoknot conformations from a lattice-based, coarse-grain tertiary structure model.
\textit{J. Chem. Phys.}, \textbf{132}, 095101.


\bibitem{oxrna}
Sulc, P., Romano, F., Ouldridge, T. E., Doye, J. P. K. and Louis, A. A. (2014)
A nucleotide-level coarse-grained model of RNA.
\textit{J. Chem. Phys.}, \textbf{140}, 235102.


\bibitem{escore}
Bottaro, S., Di Palma, F. and Bussi, G. (2014)
The role of nucleobase interactions in RNA structure and dynamics.
\textit{Nucl. Acids Res.}, \textbf{42}, 13306--13314.

\bibitem{notapaper}
Bottaro, S., Di Palma, F. and Bussi, G. (2015)
Towards de novo RNA 3D structure prediction.
\textit{RNA and Disease}, \textbf{2}, e544

\bibitem{farfar}
Das, R., Karanicolas, J. and Baker, D. (2010).
Atomic accuracy in predicting and designing  noncanonical RNA structure.
\textit{Nat. Methods}, \textbf{7}, 291--294.

\bibitem{non-red-list}
Leontis, N. B. and Zirbel, C. L. (2012). 
Nonredundant 3D Structure Datasets for RNA Knowledge Extraction and Benchmarking.
In
Leontis, N. and Westhof, E. (Eds.), 
\textit{RNA 3D structure analysis and prediction},
Springer Berlin Heidelberg,
Vol 27, pp. 281--298.


\bibitem{fr3d}
Sarver, M., Zirbel, C. L., Stombaugh, J. , Mokdad, A. and Leontis, N. B. (2008)
FR3D: Finding local and composite recurrent structural motifs in RNA 3D structures.
\textit{J. Math. Biol.}, \textbf{56}, 215--252.


\bibitem{olson1975}
W. K. Olson (1975)
Configurational statistics of polynucleotide chains. A single virtual bond treatment.
\textit{Macromolecules}, \textbf{8}, 272--275.

\bibitem{frenkel}
 Frenkel, D. and Smit, B. (1996),
 \textit{Understanding Molecular Simulation: from Algorithms to Applications},
 San Diego Academic Press.
  

  
\bibitem{anneal-prot}
Snow, M. E. (1991)
Powerful simulated-annealing algorithm locates global minimum of protein-folding potentials from multiple starting conformations.
\textit{J. Comput. Chem.}, \textbf{13}, 579--584.

\bibitem{parkpande}
Park, S. and Pande, V. S. (2007).
Choosing weights for simulated tempering.
\textit{Phys. Rev. E}, \textbf{76}, 016703.


\bibitem{tip3p}
Jorgensen, W. L., Chandrasekhar, J., Madura, J. D., Impey, R. W. and Klein, M. L. (1983)
Comparison of simple potential functions for simulating liquid water.
\textit{J. Chem. Phys.}, \textbf{79}, 926--935.

\bibitem{amber99}
Cornell, W. D., Cieplak, P., Bayly, C. I., Gould, I. R., Merz, K. M., Ferguson, D. M., Spellmeyer, D. C., Fox, T., Caldwell, J. W. and Kollman, P. A. (1995)
Second generation force field for the simulation of proteins, nucleic acids, and organic molecules.
\textit{J. Am. Chem. Soc.}, \textbf{117}, 5179--5197.

\bibitem{parmbsc0}         
P\'erez, A., March\'an, I., Svozil, D., Sponer, J., Cheatham, T. E., III, Laughton, C. A. and Orozco, M. (2007)
Refinement of the AMBER Force
Field for nucleic acids: improving the description of $\alpha\gamma$ conformers.
\textit{Biophys. J.}, \textbf{92}, 3817--3829.


\bibitem{chiol3}
Banas, P., Hollas, D., Zgarbov\'a, M., Jurecka, P., Orozco, M., Cheatham, T. E., III, Sponer, J. and Otyepka, M.(2010)
Performance of molecular mechanics force fields for RNA simulations: stability of UUCG and GNRA hairpins.
\textit{J. Chem. Theory Comput.}, \textbf{6}, 3836--3849.

\bibitem{counterions}
Joung, I. S. and Cheatham, T. E., III (2008)
Determination of alkali and halide monovalent ion parameters for use in explicitly solvated biomolecular simulations.
\textit{J. Phys. Chem. B}, \textbf{112}, 9020--9041.

\bibitem{gromacs1}
Hess, B., Kutzner, C., van der Spoel, D. and Lindahl, E. (2008)
Gromacs 4: Algorithms for Highly Efficient, Load-Balanced, and Scalable Molecular Simulation.
\textit{ J. Chem. Theory Comput.}, \textbf{4}, 435-- 447.

%
%
%
%

%
%
%
%

\bibitem{plumed}
Tribello, G. A., Bonomi, M., Branduardi, D., Camilloni, C. and Bussi, G. (2014)
PLUMED2: New feathers for an old bird.
\textit{Comp. Phys. Comm.}, \textbf{185}, 604--613.

\bibitem{kabsch-rmsd}
Kabsch, W. (1978).
A discussion of the solution for the best rotation to relate two sets of vectors.
\textit{Acta Cryst.}, \textbf{A34}, 827--828.

\bibitem{inf}
Parisien, M., Cruz, J. A., Westhof, E. and Major, F. (2009).
New metrics for comparing and assessing discrepancies between RNA 3D structures and models.
\textit{RNA}, \textbf{15}, 1875--1885.


\bibitem{molprob}
Chen, V. B., Arendall, W. B., III, Headd, J. J., Keedy, D. A., Immormino, R. M., Kapral, G. J., Murray, L. W., Richardson, J. S. and Richardson, D. C. (2010)
MolProbity: all-atom structure validation for macromolecular crystallography.
\textit{Acta Crystallogr D. Biol. Crystallogr.}, \textbf{66}, 12--21.

\bibitem{bottaro2017}
Bottaro, S. and Lindorff-Larsen, K. (2017)
Mapping the universe of RNA tetraloop folds.
\textit{Biophys. J.} {\bf 113}, 257--267.
 
\bibitem{skolnick2000}
Skolnick, J., Kolinski, A. and Ortiz, A. (2000).
Derivation of protein-specific pair potentials based on weak sequence fragment similarity.
\textit{Proteins}, \textbf{38}, 3--16.

\bibitem{pyle-pstorsions}
Wadley, L. M., Keating, K. S., Duarte, C. M. and Pyle, A. M. (2007)
Evaluating and learning from RNA pseudotorsional space: Quantitative validation of a reduced representation for RNA structure.
\textit{J. Mol. Biol.}, \textbf{372}, 942--957.


\end{thebibliography}
\end{document}